\documentclass{aa}  

\usepackage{graphicx}
\usepackage{txfonts}
\usepackage{bm}
\usepackage{amsmath}
\usepackage{siunitx}
\usepackage{xcolor}
\usepackage{hyperref}

\bibpunct{(}{)}{;}{a}{}{,} 
\usepackage{bbold}
\hypersetup{colorlinks=true,citecolor=blue,linkcolor=blue,urlcolor=blue}
\usepackage{tabularx}
\usepackage{xspace}

\usepackage{soul}
\usepackage{fontawesome}
\usepackage{ragged2e}

\usepackage{algorithm}
\usepackage{algorithmic}

\newcommand{\vilasini}[1]{{#1}}
\newcommand{\vilasinifixed}[1]{{{#1}}}

\begin{document} 

 
  \title{Generative modeling of convergence maps based on predicted one-point statistics}

   \author{Vilasini Tinnaneri Sreekanth
          \inst{1}
          \and
          Jean-Luc Starck
          \inst{1,2}
          \and
          Sandrine Codis\inst{1}
          }

   \institute{Université Paris-Saclay, Université Paris Cité, CEA, CNRS, AIM, 91191, Gif-sur-Yvette, France\\
              \email{vilasini.tinnanerisreekanth@cea.fr}
         \and
             Institutes of Computer Science and Astrophysics, Foundation for Research and Technology Hellas (FORTH), Greece       
             }

   \date{Received ?; accepted ?}

 
\abstract
   {Weak gravitational lensing is a key cosmological probe for current and future large-scale surveys. While power spectra are commonly used for analyses, they fail to capture non-Gaussian information from nonlinear structure formation, necessitating higher-order statistics and methods for efficient map generation.}
   {To develop an emulator that generates accurate convergence ($\kappa$) maps directly from an input power spectrum and wavelet $\ell_1$-norm without relying on computationally intensive simulations.}
   {We use either numerical or theoretical predictions to construct $\kappa$ maps by iteratively adjusting wavelet coefficients to match target marginal distributions and their inter-scale dependencies, incorporating higher-order statistical information.}
   {The resulting $\kappa$ maps accurately reproduce the input power spectrum and exhibit higher-order statistical properties consistent with the input predictions, providing an efficient tool for weak lensing analyses.}
   {}


   \maketitle
%

\section{Introduction}

Weak lensing, the subtle distortion of light from distant galaxies due to massive cosmic structures, is a cornerstone for mapping the Universe’s mass distribution and refining cosmological parameters. Upcoming cosmological surveys, such as the Canada-France-Hawaii Telescope Lensing Survey (CFHTLenS), Hyper Suprime-Cam (HSC), \textit{Euclid}, and \textit{LSST} \citep{2012MNRAS.427..146H, mandelbaum2017weak, laureijs2011euclid, 2019ApJ...873..111I}, aim to leverage weak lensing to address fundamental questions in cosmology \citep{Lesgourgues_2012, 2019PhRvD..99f3527L, Huterer_2010} and place tighter constraints on cosmological parameters \citep{TROXEL20151}. 

Power spectra are widely used tools for analyzing weak gravitational lensing and extracting information about the matter distribution in the Universe. However, these second-order statistics have significant limitations: they fail to capture the non-Gaussian information generated by the nonlinear evolution of structures on small scales, leading to a substantial loss of insight into gravitational and astrophysical processes. An alternative is to use
 higher-order statistics, which can better capture non-linear and non-Gaussian processes that govern cosmic structure formation. Examples include peak counts \citep{1999MNRAS.302..821K, liu2015cosmology, liu2015cosmological, 2015A&A...583A..70L, 2017A&A...599A..79P, 2020PhRvD.102j3531A}, scattering transforms \citep{2021A&A...645L..11A, Cheng2021}, Minkowski functionals \citep{2012PhRvD..85j3513K, 2020A&A...633A..71P}, higher-order moments \citep{petri2016mocking, 2018A&A...619A..38P, 2020MNRAS.498.4060G}, wavelet $\ell_1$-norms \citep{2021A&A...645L..11A, 2024A&A...691A..80S}, one-point Probability Density Functions (PDFs) \citep{bernardeau2001construction, liu2019constraining, 2021MNRAS.503.5204B} and neural networks \citep{zoltan_Ribli_2019, Fluri_2018_convergence, Fluri_2019lensing, sharma2024comparative,2023A&A...679A..61L}.
 These statistics enhance cosmological inference by capturing complex correlations and reducing degeneracies between cosmological parameters. 
 
 Accurately modeling these statistics, however, requires numerical simulations, as analytical approaches often break down in nonlinear regimes. 
 Simulations are indeed the only practical means of capturing the complex and nonlinear evolution of cosmological structures while incorporating the effects of baryons, biases, and astrophysical processes. They also enable the connection between theoretical predictions and real observations, accounting for systematic effects and uncertainties. Cosmological inference based on simulations is thus an essential step to maximize the information extracted from modern surveys and to test models of the Universe with precision. 
 This has motivated different interesting new strategies in the last years for cosmological inferences, such as Forward Modeling \citep{Alsing16, Boehm17, Remy2022, 2023A&A...679A..61L}, Field-level inference \citep{2023arXiv230404785P}  or Simulation-based inference (SBI, e.g, \cite{2024arXiv240510881G, 2021MNRAS.501..954J, papamakarios2017advances, 2020JCAP...06..050F, 2018MNRAS.477.2874A, 2020JCAP...09..048T}). 
 In the forward modeling approach, simulations are run to generate synthetic data based on a specific model and cosmological parameters. These synthetic datasets are then compared to real observational data. The comparison is often made using summary statistics (like power spectra, correlation functions, or higher-order statistics), which are easier to handle computationally and can summarize the key features of the data.
 
The goal is, therefore, not to recover the full, detailed matter density field but rather to understand how theoretical models (with specific parameters) translate into measurable quantities that can be compared to observations. Field-level inference works directly with the raw data rather than relying on simplified summaries and can, therefore, recover the full matter density field. Both forward modeling and field-level inference involve on-the-fly simulations.  In SBI, simulations are usually pre-generated based on a range of cosmological models with different parameters (e.g., matter density, dark energy, etc.). These simulations provide synthetic datasets that reflect different possible configurations of the Universe according to the models being tested.

While on-the-fly simulations offer flexibility and adaptability for modeling complex systems in real-time, their major limitations include high computational costs, potential loss of resolution and accuracy, challenges in exploring large parameter spaces efficiently, and difficulties in managing uncertainties. For SBI, its effectiveness is limited by the range and accuracy of the simulations it uses. If the simulations are not comprehensive or do not include all relevant physical processes, the inferences made could be biased or inaccurate.  Also as the number of parameters increases, the complexity of generating and comparing simulations grows significantly (i.e. curse of dimensionality). Therefore SBI  limitations are mainly due to the high computational costs, the dependence on the quality of simulations, and challenges in exploring large parameter spaces. 

So, these new ideas are very attractive but still raise challenges, especially for the analysis of large stage-IV surveys. Emulators can serve as a middle ground between traditional methods based on second-order statistics and these recent simulation-intensive techniques. A huge variety of emulators exist in the literature, \citep{2016MNRAS.463.2273F,2016MNRAS.459.2327I,2013JCAP...06..036T,2014MNRAS.437.2594W,2014MNRAS.439L..21K,2002MNRAS.329..629S}. Several approaches are grounded in analytical prescriptions that model the density field within redshift shells \citep{2023OJAp....6E..11T} or employ inverse-Gaussianization techniques \citep{2016PhRvD..94h3520Y}.  In recent years, several innovative methods have emerged, leveraging machine learning tools such as Generative Adversarial Networks (GANs) \citep{2024arXiv240605867B, 2018ComAC...5....4R} or normalizing flows \citep{2022MNRAS.516.2363D, 2023mla..confE..10D}. They are trained using a set of pre-existing simulations.  Nevertheless, a major limitation is the requirement for a large number of simulations for training, which constrains these methods to the parameter space in which they have been trained.

Another method for generating simulations of weak lensing data relies on assuming a prior on the distribution of the convergence maps. Indeed,  the convergence map is often described as a lognormal random field \citep{2011A&A...536A..85H, 2016MNRAS.459.3693X}. 
This can be very helpful for making more accurate inferences while reducing the reliance on full simulations, especially when computational resources are limited. 

In this paper, we introduce a new way of generating weak lensing convergence maps that can be obtained directly from the theory. This paper is organised as follows: in section \ref{sec:state of the art}, we give a brief review of the current state of the weak lensing emulators in place and delve a bit into the shifted log-normal models that are being used widely. In section \ref{sec:modelling weak lensing}, we recap the weak lensing model, followed by the introduction of the algorithm in detail in section \ref{sec:emulator algorithm}, \vilasini{with the full algorithmic procedure outlined in Appendix \ref{app:projections}}. Continuing to section \ref{sec:results}, we show the results we have from our algorithm and proceed to show the bench-marking in section \ref{sec:benchmarking}, before finally concluding in section \ref{sec:conclusions}. \vilasini{Appendix \ref{app:HOS} provides complementary validation tests based on other higher-order statistics, namely one-point PDFs and peak counts, which were omitted from the main text for brevity, and shows that these statistics also match remarkably well within the statistical errors of the numerical experiments.}

\section{Weak lensing convergence map and the log-normal model}
\label{sec:state of the art}

\subsection{Convergence Maps}
\label{sec convergence maps}

The convergence \(\kappa(\bm{\vartheta})\) represents the projection of the density field along the line of sight, weighted by a lensing kernel involving comoving distances \citep{1999ARA&A..37..127M}:
\begin{equation}
    \kappa(\bm{\vartheta}) = \int_0^{\chi_s} {\rm d}\chi \, \omega(\chi,\chi_s) \delta(\chi,\mathcal{D}\bm{\vartheta}),
    \label{def-convergence}    
\end{equation}
where \(\chi\) is the comoving radial distance, \(\chi_s\) is the radial distance to the source, and \(f_K(\chi)\) is the comoving angular distance:
\begin{equation}
    f_K(\chi) = \left\{
    \begin{aligned}
    & \frac{\sin (\sqrt{K} \chi)}{\sqrt{K}} \quad \text{for } K > 0, \\
    & \chi \quad \text{for } K = 0, \\
    & \frac{\sinh (\sqrt{-K} \chi)}{\sqrt{-K}} \quad \text{for } K < 0,
    \end{aligned}
    \right.
\end{equation}
with \(K\) the spatial curvature. The lensing kernel \(\omega(\chi, \chi_s)\) is defined as:
\begin{equation}
    \omega(\chi, \chi_s) = \frac{3\,\Omega_m\,H_0^2}{2\,c^2} \, \frac{f_K(\chi)\,f_K(\chi_s - \chi)}{f_K(\chi_s)} \, (1 + z(\chi)),
    \label{eq: lensing kernel}
\end{equation}
where \(c\) is the speed of light, \(\Omega_m\) is the matter density parameter, and \(H_0\) is the Hubble constant at \(z = 0\).

The observed shear field \(\gamma(\theta)\) relates to the convergence \(\kappa\) via mass inversion \citep{1993ApJ...404..441K,starck2006,2015A&A...581A.101M}. 

\subsection{Angular $\mathrm{Cl_s}$}
Second-order statistics, such as the shear two-point correlation functions \(\xi_{\pm}(\theta)\) and their Fourier-space counterpart, the angular power spectrum \(C_{\ell}\), play a crucial role in extracting cosmological information from weak lensing surveys. These statistics describe the spatial distribution of the shear field and provide a window into the underlying matter distribution and its evolution.

Within the framework of the Limber approximation, the angular power spectrum of the convergence field for a specific tomographic bin is expressed as \citep{2015RPPh...78h6901K}:
\begin{equation}
    C_{\kappa}(\ell) = \frac{9 \Omega_m^2 H_0^4}{4 c^4}
    \int_0^{\chi_{\text{lim}}} \text{d}\chi \,
    \frac{g^2(\chi)}{a^2(\chi)} P_{\delta}
    \left( k = \frac{\ell}{f_K(\chi)}, \chi \right),
\end{equation}
where \(P_{\delta}\) represents the matter power spectrum of the density contrast, \(\chi\) denotes the comoving distance, and \(g(\chi)\) is the lensing kernel that quantifies lensing efficiency as a function of distance. The term \(a(\chi)\) refers to the scale factor at distance \(\chi\), while \(f_K(\chi)\) is the comoving angular diameter distance, determined by the spatial curvature of the universe.

This formulation enables the computation of the convergence power spectrum, providing insights into structure growth and the geometry of the universe by linking weak lensing observables to the underlying matter distribution.

\subsection{Shifted Log-normal Model}

\begin{figure*}
    \centering
    \includegraphics[width=16cm]{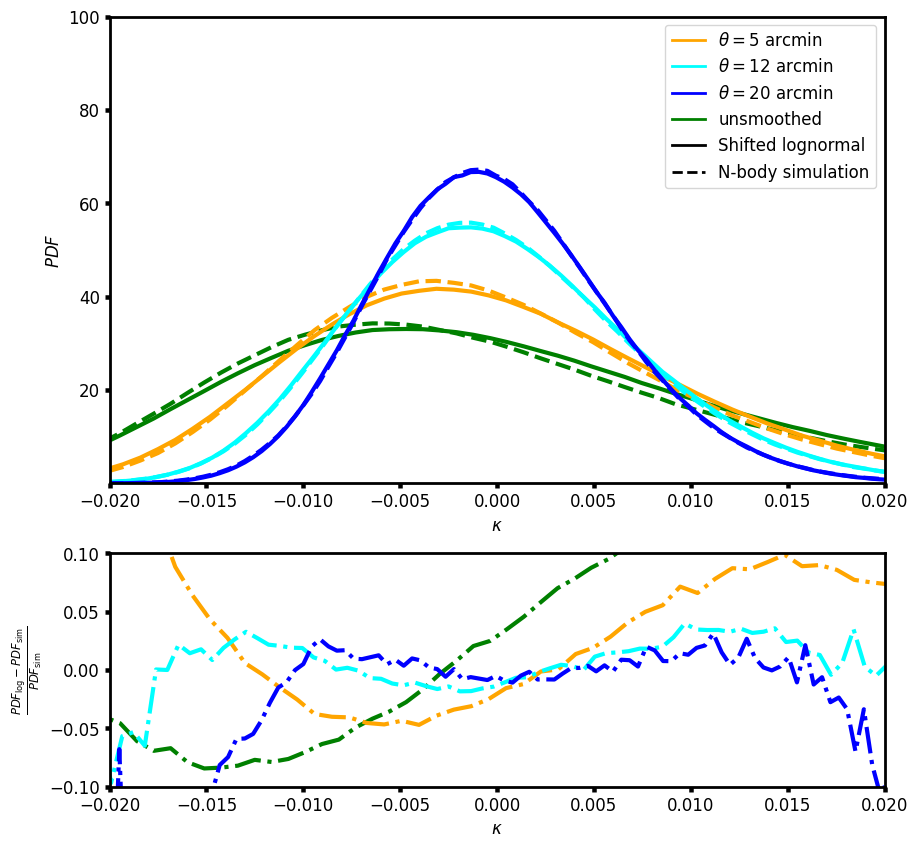}
    \caption{Comparison of the one-point PDF from N-body simulations and from maps generated by shifted log-normal modeling. Top panel: Distributions of the shifted log-normally generated maps (solid lines) and N-body simulations (dashed lines) are shown for the unsmoothed case (in green)  and at a smoothing angle of $ \theta = 5 \, \mathrm{arcmin} $, $ \theta = 12 \, \mathrm{arcmin} $ and at $ \theta = 20 \, \mathrm{arcmin} $ as labelled. Bottom panel: \vilasinifixed{Relative residuals of the log-normal PDF with respect to the N-body simulation.}}
    \label{fig: lognormal demo}
\end{figure*}

In cosmology, random fields like the matter density contrast or lensing convergence are also described statistically using probability density functions (PDFs), which define how likely different values of the field are to occur. For some fields, the PDF is well-approximated by a Gaussian distribution.
However, many cosmological fields deviate significantly from this Gaussian behaviour. For instance, the matter density contrast and lensing convergence have hard lower limits (e.g., density cannot drop below zero relative to the average) and exhibit skewed shapes with heavy tails, indicating a higher likelihood of extreme values compared to a Gaussian distribution. In such cases, a shifted lognormal \citep{1991MNRAS.248....1C,2016MNRAS.459.3693X,2023OJAp....6E..11T} distribution offers a better model, and has emerged as a practical alternative to full N-body simulations. This distribution can account for the asymmetry, hard limits, and heavy tails observed in the data, providing a more realistic representation of these fields' statistical behaviour.

This approach is particularly effective in capturing the non-Gaussian distribution of matter, while accounting for negative convergence values observed in specific scenarios. By shifting the lognormal distribution, the model extends its range to include regions with negative convergence, thereby improving the realism and accuracy of density fluctuation simulations. This enhancement is particularly valuable for representing under-dense regions while preserving the nonlinear clustering effects of large-scale structures.

Under moderate noise levels and typical accuracy requirements, the lognormal model provides reliable covariance matrices and bispectra \citep{2022PhRvD.105l3527H}. However, the skewness of the lognormal model alone is insufficient to fully capture the nonlinear features induced by gravitational evolution, leading to inaccuracies in higher-order statistics \citep{2023MNRAS.520..668P}.

Given a set of variables \( \mathrm{K_{g_i}} \) following a multivariate Gaussian distribution with mean vector \( \mu_i \) and covariance matrix \( \xi_{ij}^{g} \), the shifted lognormal transformation is expressed as:

\begin{equation}
    \mathrm{K_{logn_i}} = e^{\mathrm{K_{g_i}}} - \lambda_i,
\end{equation}

where \( \lambda_i \) depends on the cosmology and is computed as a function of the expected mean, variance, and skewness of the distribution. This dependence makes the model more flexible than a simple Gaussian approximation. Several approaches are used to fine-tune the shift parameter that is used in a lognormal model. The common approach is to focus on one scale of interest, and obtain a shift parameter such that the lognormal model has the correct skewness as that of the simulation, i.e it is calibrated against the PDF of the field at this particular scale, and that the shift parameter is cosmology and scale dependent. This means that fixing the shift parameter at one scale does not guarantee the skewness to be exact across all the other scales. Since convergence is not strictly a log-normal variable, different methods for determining \( \lambda_i \) yield different values, potentially compromising the model's accuracy at other scales. An added complexity arises when modeling the matter density field as a log-normal field and integrating along the line of sight to derive the convergence map, further complicating the estimation of \( \lambda_i \). For further details, refer to \citet{2016MNRAS.459.3693X}.

This log-normal transformation modifies the correlation function as follows \citep{2016MNRAS.459.3693X}:

\begin{equation}
    \xi_{ij}^{\mathrm{logn}} = \lambda_i \left(e^{\xi_{ij}^g} - 1\right).
    \label{eq:corr_func}
\end{equation}

Hence, to generate a log-normal field with the desired correlation function, the Gaussian field to which the local transformation is applied must first be constructed with a corrected correlation function.


As expected, Equation \ref{eq:corr_func} shows that the correlation function -- and hence the variance -- is independent of the shift parameter, which is often calibrated so as to reproduce the expected skewness of the log-normal field. Note that the correlation function is related to the power spectrum via:
\begin{equation}
    \mathrm{C^{ij}_{logn} (\ell)} = 2 \pi \int_{0}^{\pi} d\theta \sin \theta \, P_{\ell}(\cos\theta) \, \xi^{g}_{ij}(\theta),
\end{equation}
where \( P_{\ell} \) represents the Legendre polynomial of order \( \ell \).

To illustrate the limitations inherent to the (shifted) log-normal model, \vilasinifixed{We compare the convergence map generated using a log-normal model to one derived from an $N$-body simulation. For this comparison, we utilize the publicly available Takahashi simulation suite \citep{Takahashi_2017}, which provides full-sky convergence maps at a fixed cosmology \footnote{\url{http://cosmo.phys.hirosaki-u.ac.jp/takahasi/allsky_raytracing/}}. These maps are available at two grid resolutions: $4096$ and $8192$, and include data spanning the full sky at intervals of $150 \, h^{-1}$ comoving radial distance, covering redshifts from $z = 0.05$ to $5.3$. The smoothing of the convergence maps is performed through a direct convolution of the map with a wavelet filter at specified angular scales. For this analysis, we use the convergence map at $\text{nside} = 4096$, which is subsequently downgraded to $\text{nside} = 1024$. A corresponding log-normal map with $\text{nside} = 1024$ and a maximum $\ell$-mode of $\ell_{\text{max}} = 3 \, \text{nside} + 1$ is then generated for comparison.
} Figure \ref{fig: lognormal demo} compares the distribution of a log-normally generated convergence map with N-body simulations. The shift parameter is optimised to match the distribution of the N-body simulation at a smoothing scale of \( \theta = 20 \, \mathrm{arcmin} \). However, the figure illustrates that this configuration fails to hold at other scales or for the full map.

This limitation underscores the necessity of developing a more flexible emulator capable of functioning consistently across all scales, while effectively capturing cross-scale correlations. Such advancements are crucial for robust multi-scale analyses, enabling the extraction of nonlinear information essential for cosmological inference.

\section{Emulator based on the wavelet l1-norm theoretical prediction}
\label{sec:modelling weak lensing}

Recent advancements in theoretical modeling have focused on higher-order statistics such as the probability distribution function (PDF) \citep{2021MNRAS.503.5204B, 2021MNRAS.505.2886B} and the wavelet $\ell_1$-norm \citep{2024A&A...691A..80S}, leveraging Large Deviation Theory. These models have demonstrated the capability to provide accurate predictions of the one-point statistics (PDF and wavelet $\ell_1$-norm) of the density and convergence field up to mildly non-linear scales Indeed, it has been shown that for the wavelet $\ell_1$-norm of the convergence maps, prediction can be obtained at sub percent accuracy for a $\kappa$ map smoothed up to $ \theta = 15 \, \mathrm{arcmin} $, for a source redshift $z_s \approx 2$.  

Our goal is hence to build an emulator based on the theoretical prediction of the $\ell_1$-norm. In the following section, we describe our algorithm called (GOLCONDA  : Generative mOdeLing of Convergence maps based ON predicteD one-point stAtistics) in detail and present our results.

\subsection{Wavelet $\ell_1$-norm }

The wavelet transform is a powerful tool in the analysis of astronomical images due to its ability to decompose data into components across different scales. This multi-scale decomposition is particularly effective for studying the complex, hierarchical structures often found in astronomical data. We refer the readers to \citet{book} for more details. 

In this work, we employ a wavelet transform derived from using a top-hat filter as the scaling function as is explained in \citep{2024A&A...691A..80S}.
An image \(c_0\) can then be decomposed as the sum of all wavelet scales and a coarse resolution image \(c_J\):
\begin{equation}\label{wav_des}
    c_0(x,y) = c_J(x,y) + \sum_{j=1}^{J_{\text{max}}} w_j(x,y),
\end{equation}
where \(J_{\text{max}}\) denotes the maximum number of scales, and \(w_j\) represents wavelet images that capture the details of the original image at dyadic scales $j$.
%
Each set of wavelet coefficients \(w_j\) is generated by convolving the input map with the respective wavelet kernel.

The wavelet $\ell_1$-norm is then defined as
\begin{equation}
    \ell_1^{\theta_j}[i] = \sum_{k=1}^{\# coef(S_{\theta_j,i})} |S_{\theta_j,i}[k]|,
    \label{eq: l1norm}
\end{equation}
where the set of coefficients at scale ${\theta_j}$ and amplitude bin $i$, 
$S_{\theta_j,i} = \left\{w_{\theta_j}~  / ~~  w_{\theta_j}(x,y)  \in [ B_i,  B_{i+1} ] \right\}$, 
depicts the wavelet coefficients $w_{\theta_j}$ having an amplitude within the bin $ [B_i,  B_{i+1}]$, and the pixel indices are given by $(x,y)$.
In other words, for each bin, the number of pixels $k$, that fall in the bin is collected and the absolute values of these pixels are summed to obtain the $\ell_1$-norm at this bin $i$. 

This definition allows for capturing data through the absolute values of all pixels within the map rather than limiting the analysis to identifying local minima or maxima. It offers the benefit of a multiscale approach and was shown to be particularly robust compared to other approaches \citep{huber1987place, gine2003bm}.

\subsection{Theoretical prediction of the wavelet $\ell_1$-norm}

In \citet{2024A&A...691A..80S}, a theoretical framework was developed to predict the wavelet \( \ell_1 \)-norm, denoted as \( \ell_1^{\theta_j}[i] \), for weak lensing convergence maps. This framework leverages Large Deviation Theory (LDT), a statistical method to describe rare events, to derive the one-point probability density function (PDF) of the convergence field. In this framework, the wavelet \( \ell_1 \)-norm for a specific bin \( i \) is expressed as:
\begin{equation}
    \ell_1^{\theta_j}[i] = |S_{\theta_j,i}| \, P(S_{\theta_j,i}) \, \mathcal{N},
    \label{eq:l1norm_pdf}
\end{equation}
where \( |S_{\theta_j,i}| \) is the bin value (absolute value of the wavelet coefficient),
   \( P(S_{\theta_j,i}) \) is the normalized PDF of the wavelet coefficients at scale \( \theta_j \),
  and \( \mathcal{N} \) is a factor accounting for the normalization of the predicted PDF.

The theoretical predictions for \( \ell_1^{\theta_j}[i] \) were validated against results from N-body simulations, demonstrating agreement at the percent level. This approach provides an efficient and accurate alternative to computationally expensive simulations for analyzing non-Gaussian features in weak lensing data. For further details, we refer the reader to \citet{2024A&A...691A..80S}.

Given the simplicity of predicting the wavelet $\ell_1$-norm for any specified cosmology, the objective is to utilize this prediction alongside the power spectrum as a constraint to generate a weak lensing $\kappa$-map. The following section explores how these inputs can be effectively employed to produce a map that satisfies the desired statistical properties.

\subsection{Optimisation problem}

Our goal is to emulate a map with the statistical properties predicted by theoretical cosmological models for a given set of parameters. As the GLASS emulator, we want to emulate a map $\kappa$ which has the right theoretical power spectrum $P_t$,  i.e. $\mathcal{P}(\kappa) = P_t$, where  $\mathcal{P}$ is the operator computing the radial power spectrum of $\kappa$.  We also want the same map to have the right distribution in the wavelet space, as predicted by LDT. 
Noting, $W_\kappa = \left\{w_1, \dots, w_{J-1}, c_{J-1}\right\}$  the wavelet transform of $\kappa$ using $J$ scales,  i.e. W = $\Phi^t \kappa$, and $L_{1,t} =  \left\{ L_{1,t}^{\theta_1} ,  \dots, L_{1,t}^{\theta_2} , \dot,  \dots, L_{1,t}^{\theta_J} \right\}$ the theoretical prediction in the different scales, 
each wavelet scale  $w_j$ should have the correct $l_1$-norm distribution  $ \ell_1^{\theta_j,i} $, 
i.e.  $\mathcal{L}_1^{\theta_j}(w_j) =  L_{1,t}^{\theta_j} $, 
where $\mathcal{L}_1^{\theta_j}$ is the operator calculating the $\ell_1$-norm at scale $j$ following Eq.~\ref{eq: l1norm}. 

\vilasini{Specifically, we aim to ensure that:
\begin{itemize}
    \item The map has the correct theoretical power spectrum $P_t$: 
    \begin{equation}
    \mathcal{P}(\kappa) = P_t,
    \label{eq:power_constraint}
    \end{equation}
    where $\mathcal{P}$ is the operator computing the radial power spectrum.
    \item The wavelet coefficients of the map have the correct $\ell_1$-norm distribution at each scale:
    \begin{equation}
    \forall j \in \{1, \dots, J-1\}, \quad \mathcal{L}_1^{\theta_j}(w_j) = L_{1,t}^{\theta_j},
    \label{eq:l1_constraint}
    \end{equation}
    where $W_\kappa = \{w_1, \dots, w_{J-1}, c_{J-1}\} = \Phi^t \kappa$ is the wavelet transform, and $\mathcal{L}_1^{\theta_j}$ is the operator computing the $\ell_1$-norm at scale $j$.
\end{itemize}
}




\vilasini{We introduce two projection operators associated with these constraints:
\begin{itemize}
    \item $\mathcal{C}_f$: projection onto the space of maps satisfying the power spectrum constraint (Eq.~\ref{eq:power_constraint}).
    \item $\mathcal{C}_w$: projection onto the space of wavelet coefficients satisfying the $\ell_1$-norm constraints (Eq.~\ref{eq:l1_constraint}).
\end{itemize}}

The problem is then formulated as a constrained optimisation:
\begin{equation}
\min_{\kappa, W} \; \| \kappa - \Phi W \|^2 
\quad \text{s.t.} \quad 
\mathcal{C}_f(\kappa) = \kappa, \quad 
\mathcal{C}_w(W) = W.
\label{eq:minimization}
\end{equation}

\vilasini{This minimisation ensures that $\kappa$ and $W$ are consistent through the wavelet transform, while satisfying the power spectrum and wavelet norm constraints. Because these constraints correspond to different domains (Fourier and wavelet space), a joint solution is not trivial and requires an iterative strategy. The Generalised Forward-Backward (GFB) algorithm \citep{2015arXiv150303703L} is well-suited for problems involving multiple non-smooth convex constraints. It enables us to decompose the optimisation into simpler sub-problems associated with each constraint and alternate between them efficiently while ensuring convergence. In our case, the algorithm alternates between applying the projections $\mathcal{C}_f$ and $\mathcal{C}_w$, while enforcing consistency between $\kappa$ and $W$.}

Using an alternate minimisation approach, we have:
\begin{itemize}
    \item Assuming $W$ is known, minimise:
     \begin{eqnarray}
 \min_{\kappa}    \parallel  \kappa - \Phi W \parallel^2  ~~~s.t.~~~   
   {\cal{C}}_f (\kappa) = \kappa     
 \end{eqnarray}
     \item Assuming $\kappa$ is known, minimise:
 \begin{eqnarray}
 \min_{W}    \parallel  \kappa - \Phi W \parallel^2  ~~~s.t.~~~   
    {\cal{C}}_w (W) =  W  
 \end{eqnarray}
\end{itemize}


\vilasini{The GFB algorithm applied to this setup yields the following iterative update scheme:}
\begin{enumerate}
\item $z_1 = {\cal{C}}_p(\Phi W^{n})$
\item $z_2 = {\cal{C}}_w ( W^n +  \Phi^t ( \kappa^n - \Phi W^{n})) $
\item $\kappa^{n+1} = \frac{1}{2} ( z_1 +  \Phi z_2)$
\item $W^{n+1} =  \frac{1}{2} ( \Phi^t z_1 +  z_2)$
\end{enumerate}

Note that any additional summary statistics with theoretical prediction could be incorporated in a straightforward way, adding an additional constraint on $\kappa$.

The ${\cal{C}}_p$ operator consists in i) taking the Fourier transform $\hat{\kappa}$ of $\kappa$, ii) computes its radial power spectrum $P_{\kappa}$, iii) multiply the Fourier coefficients of $\hat{\kappa}$ by $\sqrt{\frac{P_t [k] }{P_{\kappa} [k] }}$ (i.e. $z [u,v] = \sqrt{\frac{P_t [k] }{P_{\kappa} [k] }}   \hat{\kappa}  [u,v] $, with $k=\sqrt{u^2+v^2}$), and iv) take the inverse Fourier transform of $z$. Similarly, the  ${\cal{C}}_w$ operator independently adjusts the  $\ell_1$-norm of all wavelet scales. \vilasini{The projection operators $\mathcal{C}_f$ and $\mathcal{C}_w$ are described in detail in Appendix~\ref{app:projections}.}

When $\Phi$ is orthogonal, we have $\Phi^t \Phi = Id $, and the step 2 of the previous equation becomes:
     \begin{eqnarray}
z_2 & = & {\cal{C}}_w ( W^n +  \Phi^t \kappa^n - \Phi^t \Phi W^{n}) \nonumber \\ 
    & = & {\cal{C}}_w ( \Phi^t \kappa^n )
 \end{eqnarray}
The method becomes therefore extremely simple, just applying iteratively the two steps:
     \begin{eqnarray}
\kappa^{n+1} & = &  \frac{1}{2} ( {\cal{C}}_p(\Phi W^{n}) +  \Phi {\cal{C}}_w ( \Phi^t \kappa^n ))  \nonumber \\
& = &  \frac{1}{2} ( {\cal{C}}_p(\kappa^n) +  \Phi {\cal{C}}_w ( W^{n} ))  \nonumber \\ 
W^{n+1} & = &  \frac{1}{2} (  \Phi^t  {\cal{C}}_p(\Phi W^{n}) +  {\cal{C}}_w ( \Phi^t \kappa^n ))  \nonumber \\
& = & \Phi^t \kappa^{n+1}, 
 \end{eqnarray}
and we can fix $\kappa^{0}$ to a Gaussian or log-normal realisation, and $W^{0} = \Phi^t \kappa^{0}$. 
In the most general cases, the wavelet transform is a frame and is not necessary orthogonal, so we have $\Phi \Phi^t  = Id $ (due to the fact that we have a transform with exact reconstruction),
but we don't always have $\Phi^t \Phi = Id $. It remains however a reasonable approximation that allows us to speed up the processing. 

\subsection{Algorithm}
\label{sec:emulator algorithm}
The algorithm iteratively adjusts an image to satisfy two main constraints: matching a target power spectrum through Fourier amplitude correction and achieving desired \(\ell_1\)-norm wavelet coefficients. This iterative process ensures that the final image accurately represents both the specified statistical properties and the target power spectrum.

To do so, the algorithm begins with an initial image, a realisation of a 2D Gaussian random field in our case. It proceeds by iteratively refining the image through a two-step process. First, an \(\ell_1\)-norm-based wavelet coefficient adjustment (WCA) is applied to align the image's wavelet coefficients with the desired properties. Second, the Fourier amplitudes are corrected to ensure the image matches a specified target power spectrum. After each iteration, the refined image is obtained as a weighted average of the two corrected images. This iterative process continues until a convergence criterion, typically based on the relative root mean square (RMS) error between iterations, is satisfied or a predefined maximum number of iterations is reached. The algorithm involved here is given in algorithm \ref{algo:iterative_correction}.

\begin{algorithm}
\caption{Iterative Map Generation with Power Spectrum and Wavelet Constraints}
\label{algo:iterative_correction}

\textbf{Input:} 
\begin{itemize}
    \item Initial map $\kappa^0$
    \item Maximum iterations $max\_iter$
    \item Tolerance $\epsilon$
\end{itemize}

\textbf{Output:} Final map $\kappa$

\begin{algorithmic}[1]
    \STATE Initialize $\kappa \gets \kappa^0$
    \STATE Compute initial wavelet coefficients: $W^0 \gets \Phi^t \kappa^0$
    \FOR{$n = 1$ to $max\_iter$}
        \STATE Apply power spectrum constraint: $z_1 \gets \mathcal{C}_p(\Phi W^n)$
        \STATE Apply wavelet $\ell_1$-norm constraint: $z_2 \gets \mathcal{C}_w(W^n + \Phi^t (\kappa^n - \Phi W^n))$
        \STATE Update the map: $\kappa^{n+1} \gets \frac{1}{2} (z_1 + \Phi z_2)$
        \STATE Update wavelet coefficients: $W^{n+1} \gets \frac{1}{2} (\Phi^t z_1 + z_2)$
        \STATE Compute error: $error \gets \text{RelativeRMS}(\kappa^{n+1}, \kappa^n)$
        \IF{$error < \epsilon$}
            \STATE Converged. Exit loop.
            \STATE \textbf{BREAK}
        \ENDIF
        \STATE Update $\kappa \gets \kappa^{n+1}$
    \ENDFOR
    \RETURN $\kappa$
\end{algorithmic}
\end{algorithm}

\section{Results and validation of the emulator}
\label{sec:results}
\label{sec: results}

\subsection{Simulations}
In this paper, to test our pipeline, we used the Scinet
LIght-Cones Simulations (SLICS) \citep{2018MNRAS.481.1337H} suite of 924 fully independent N-body simulations, that is available publicly\footnote{SLICS:https://slics.roe.ac.uk}. They are derived from a set of 1025 N-body simulations generated using the high-performance gravity solver CUBEP3M \citep{2013MNRAS.436..540H}, which computes the nonlinear evolution of $1536^3$ particles starting from the initial conditions created at z=120 using the Zeldovich approximation, in a box of length $505 h^{-1} 
$Mpc on one side. The N-body code calculates the non-linear evolution of these collisionless particles up to $z = 0$.
Using a multiple-plane technique, convergence and shear maps were generated over an area of 100 square degrees at 18 distinct source planes, using the Born approximation.

\subsection{Validation}

\begin{figure*}
    \centering
    \includegraphics[width=\linewidth]{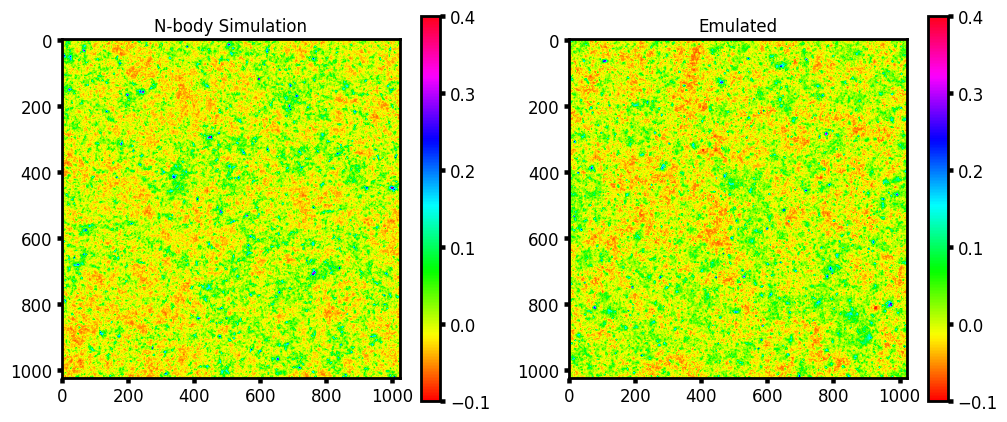}
    \caption{The target convergence map and the emulated convergence map obtained from our pipeline.\vilasinifixed{The target map is obtained for a source redshift at $z_s=2.134$. The target and the generated maps both have a resolution of 1024 pixel son each side, with an angular size of $10^2deg$. The x-axis and the y-axis gives the number of the pixels and the pixel value is given by the colorbar.} }
    \label{fig: target emulated}
\end{figure*}

To evaluate the performance of our emulator, we conducted a series of tests using a target image that was wavelet-decomposed into multiple scales, employing a top-hat filter as the scaling function. The set of equations governing the wavelet decomposition was previously defined in equation \ref{wav_des}. As mentioned previously our approach is motivated by recent theoretical advancements utilising the wavelet $\ell_1$-norm, as predicted within the LDT framework  \citep{2024A&A...691A..80S}. These predictions provide input for generating emulated maps and are primarily valid in mildly non-linear scales, which consequently constrain the resolution of the maps. To demonstrate the effectiveness of this method, we validated our algorithm against the SLICS simulation suite.

In this work, we use a target image and decompose it into a maximum of $2\log(N)$ scales, where $N$ is the number of pixels along one dimension of the map. This decomposition yields $2\log(N)$ sets of wavelet coefficients, along with a single coarse-scale image. The $\ell_1$-norm values, computed per bin for both the wavelet coefficients and the coarse-scale image, are treated as the target wavelet distributions. Simultaneously, the $\mathrm{C}_\ell$ values derived from the target image are designated as the target power spectra. Using these target distributions and power spectra as constraints, the algorithm is applied to generate emulated maps, enabling us to assess its accuracy and robustness. For validation, we generate flat-sky map of size $100 deg^2$ with 1024 pixels on each side.

Figure \ref{fig: target emulated} illustrates the original target image compared to the emulated image generated through our pipeline. The target image is derived from the N-body simulation, while the reconstructed solution is produced by our software. At first glance, no significant visual differences are observed between the target and the reconstructed images up to the phases which are different in the two cases.

During the iterative process, the relative error in the wavelet $\ell_1$-norms across different scales was assessed to evaluate the emulator's performance. This error was calculated by comparing the $\ell_1$-norm of the wavelet coefficients and the coarse-scale coefficient of the reconstructed image to those of the target image. Convergence was determined based on the stabilisation of these relative errors, indicating that the emulator had successfully approximated the target distributions.

\begin{figure}
    \centering
    \includegraphics[scale=0.5]{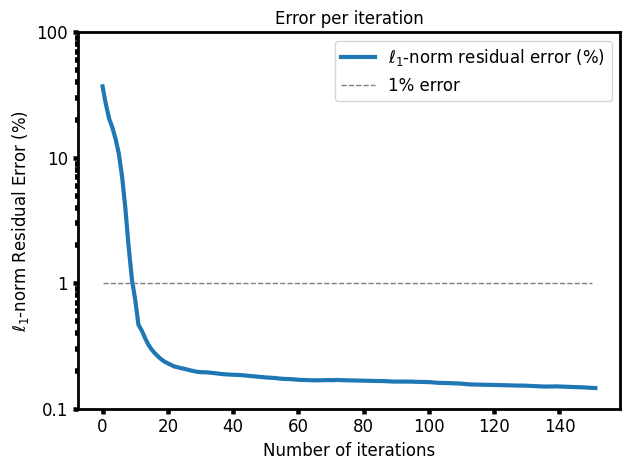}
    \caption{Evolution of the total relative error (in \%) in the \(\ell_1\)-norm of the reconstructed emulated image compared to the simulation map during the iterative process. The total relative error proves a measure of how closely the reconstructed image approximates the target simulation map. Stabilisation of this error indicates convergence of the emulator towards the target distribution.}
    \label{fig:rrmse}
\end{figure}

The number of iterations in this work is fixed at $ n_{\text{iter}} = 150 $, as multiple tests indicated good convergence at this value. This conclusion was reached after verifying that the overall relative error, defined in Equation \ref{eq: relative error}, in the $\ell_1$-norm falls below 1\%. This served as the convergence criterion to ensure that our iterative algorithm successfully converged. \vilasini{In our implementation, we set the tolerance $Err_{\ell_1} = 0.01$, corresponding to a 1\% relative error threshold. This value was chosen empirically, as we observed that both the $\ell_1$-norm and the power spectrum residuals reliably stabilised below this level, indicating convergence of the key statistical properties. While tighter thresholds could be possible, we found that $Err_{\ell_1} = 0.01$ offers a practical trade-off between computational cost and statistical accuracy. The convergence behaviour is shown in Figure~\ref{fig:rrmse}, where the relative error falls below 1\% within approximately 100 iterations.
}

\begin{equation}
    Err_{\ell_1} = \sum_{j} \left( \frac{\sum_{i} \lvert {\ell_{1}}_{\text{target}}^{i,j} - \ell_{1_emulated}^{i,j} \rvert}{\text{max}(\ell_{1_target}^j)} \right)
    \label{eq: relative error}
\end{equation}

Figure \ref{fig:rrmse} illustrates the evolution of the relative error over successive iterations. The plot highlights the error in the total $\ell_1$-norm of the emulated image compared to the target image. This result demonstrate the convergence of the algorithm and its capability to replicate the target distributions across different scales with high accuracy. From the plot we see that the algorithm converges quickly, we reach percent accuracy at around 100 iterations.

\begin{figure}
    \centering
    \includegraphics[width=\columnwidth]{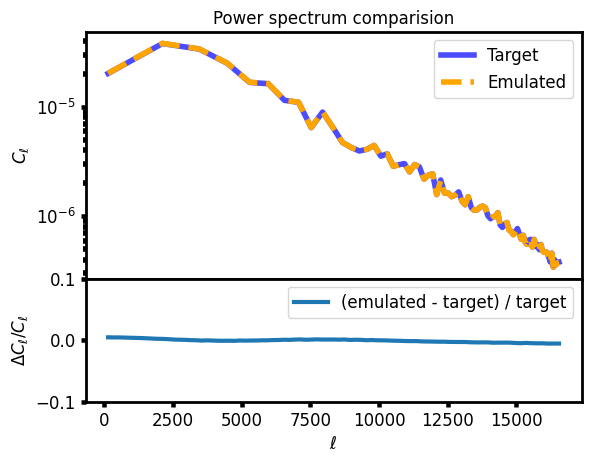}
    \caption{Power spectra (\(\mathrm{C}_\ell\)) of the target \(\kappa\) (orange solid line) map and the final emulated image (blue solid line) produced by our algorithm. The residuals (green dashed line) between the target and emulated power spectra are also shown, indicating the level of agreement between the two maps. The residual values are maintained at a sub-percent level, demonstrating the high accuracy and efficiency of the emulation process in reproducing the target power spectra.}
    \label{fig:cls}
\end{figure}

\begin{figure*}
    \centering
    \includegraphics[width=\textwidth]{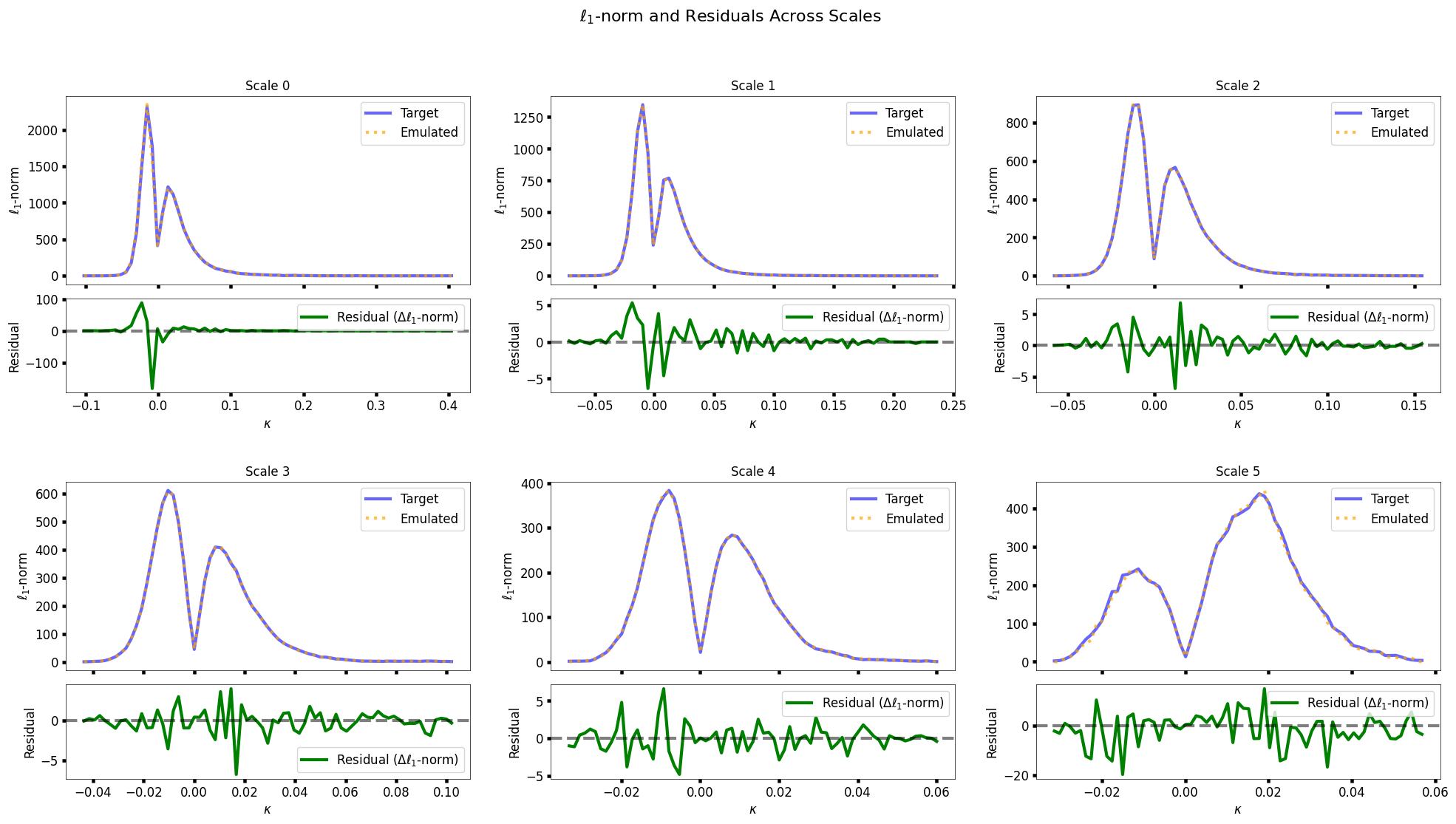}
    \caption{\vilasini{Wavelet $\ell_1$-norm at each scale and its residual between the decomposed target image (blue solid line) and the emulated image (orange dashed line) across different scales. The residuals, which are obtained as the difference between the target and the emulated wavelet $\ell_1$-norms, are also presented (green dashed line), reflecting the accuracy of the emulation process in reproducing the target distributions. The residuals consistently remain at a sub-percent level, indicating precise reconstruction and efficient performance of the algorithm.}}
    
    \label{fig: l1norm}
\end{figure*}

To conduct a more comprehensive comparison, we analyse the statistical properties of the reconstructed map, focusing on the power spectrum and the wavelet $\ell_1$-norm. These comparisons are presented in Figures \ref{fig:cls} and \ref{fig: l1norm}. 

Figure \ref{fig:cls} presents the power spectra of the target $\kappa$ map and the emulated map produced by our algorithm. The target power spectrum is derived from an N-body simulation, while the emulated power spectrum is generated using our pipeline, constrained by the input power spectrum and the wavelet $\ell_1$-norm. The figure also shows the residuals of the power spectra, calculated as the relative difference between the target and emulated spectra across all angular multipoles. These residuals remain consistently at a sub-percent level, demonstrating the algorithm's capability to replicate the target power spectrum with high precision.

In Figure \ref{fig: l1norm}, we present the wavelet $\ell_1$-norm of the decomposed image at various scales. The comparison highlights the wavelet $\ell_1$-norm values computed for the target wavelet coefficients and those obtained from the emulated wavelet coefficients at each scale. The residuals, shown alongside, quantify the differences between the target and emulated wavelet $\ell_1$-norms. Across all scales, the residuals are also at a sub-percent level, confirming the accuracy of the emulator in reproducing the target wavelet distributions.

These results emphasize the robustness of our algorithm, which successfully captures both the global and scale-dependent statistical properties of the target map. The sub-percent residuals observed in both the power spectrum and wavelet $\ell_1$-norm validate the consistency of our approach and its potential for applications in cosmological analyses where precision is critical.

\subsection{Benchmarking}
\label{sec:benchmarking}

In this section, we explore the time complexity, statistical properties, and higher-order statistics of the generated maps at multiple scales using different wavelet and smoothing filters (hence beyond the constraints used by the algorithm). This analysis provides insights into the strengths and limitations of the algorithm, highlighting its ability to emulate realistic data with minimal computational resources.

\subsubsection{Performance}
To generate the plots of this paper, our pipeline was able to generate each 500x500 map in less than 2 minutes. The figure \ref{fig:time_analysis} below presents a plot of the processing time against map size. In this study, we utilised a single CPU and one core without incorporating any parallelisation or Just-In-Time (JIT) compilation, yet it still performs comparably or better than existing methods.

\begin{figure}
 \centering
 \includegraphics[scale=0.5]{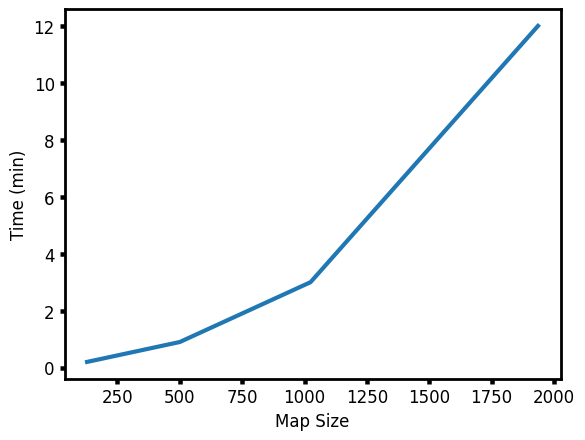}
 \caption{Time complexity of the generator. Here on the x-axis the number of pixels on each side is given. On the y-axis the time taken to generate a map for the given size is plotted}
 \label{fig:time_analysis}
\end{figure}

\subsubsection{Higher Cumulants}
\begin{figure*}
    \centering
    \includegraphics[width=\linewidth]{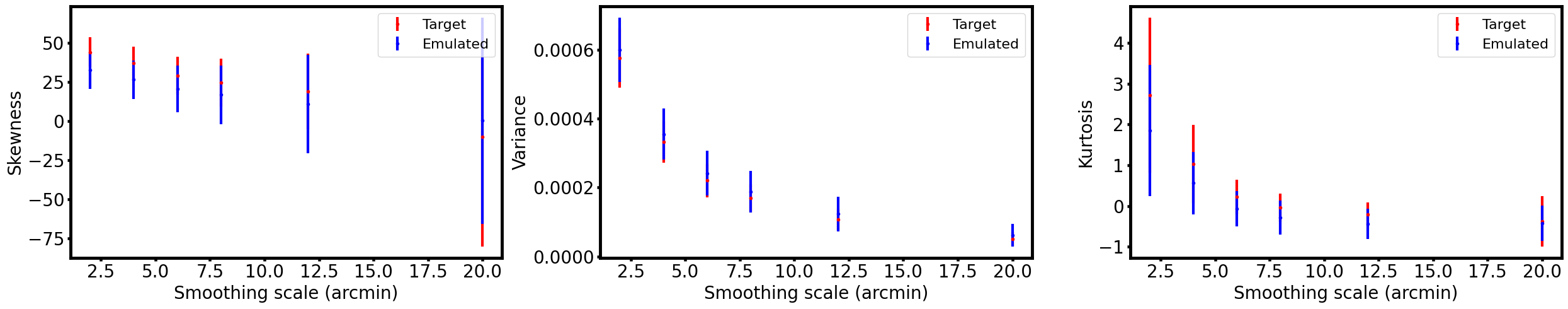}
    \caption{This plot displays the variance and skewness for the target image (in red) and the emulated image (in blue), obtained using a top-hat scaling function to derive the wavelet. The emulated image was created by an algorithm that matches the wavelet $\ell_1$-norm of the target image from the SLICS simulation. Benchmarking was conducted by smoothing both images with different radii via a top-hat filter, partitioning them into patches, and calculating skewness, variance, kurtosis, and their associated error bars \vilasini{by taking the standard deviation of the values obtained from 4 different patches of the maps}. The x-axis represents the smoothing scale in arcminutes.}
    \label{fig:variance skewness}
\end{figure*}

We further examined the statistical properties embedded in the maps, including skewness, variance, and kurtosis at various scales over the entire map. To estimate the associated error bars, we divided each map into four patches and computed the standard deviation of the moments obtained across these patches. These results are displayed in Figure \ref{fig:variance skewness}. The analysis was performed by dividing both the target and reconstructed maps into sub-patches, allowing for the computation of these statistics across the entire map and estimation of associated uncertainties.

The left panel shows the skewness, which quantifies the asymmetry of the distribution of pixel values in the maps. The skewness values and associated errors of the target (red) and reconstructed (blue) maps agree closely, demonstrating the ability of the reconstruction process to preserve the asymmetry of the target map.

The middle panel presents the variance, which measures the dispersion of pixel values. As expected, the variance decreases with increasing smoothing scale due to the suppression of small-scale fluctuations during smoothing. The reconstructed map successfully reproduces the target map’s variance at all scales, with a good overlap between the red and blue points within the error bars.

The right panel shows the kurtosis, which captures the degree of "tailedness" in the distribution of pixel values. The kurtosis generally decreases with increasing smoothing scale, consistent with the suppression of higher-order moments in smoothed data. The reconstructed map closely matches the target map's kurtosis, with only minor deviations at some scales.

Overall, the results indicate that the reconstruction process preserves the higher-order statistical properties of the target map as evidenced by the close agreement across skewness, variance, and kurtosis.

\vilasini{Appendix~\ref{app:HOS} presents some additional tests omitted other higher-order statistics namely PDFs and peak counts and shows that the incidentally match very well (within the statistical errors of the numerical experiment).}

\subsection{Extension to full-sky maps generation}

Extending the algorithm to full-sky maps is a straightforward process, as the methodology remains applicable in this setting. To do that, we use \citep{Zonca2019, 2005ApJ...622..759G}.  This extension has already been implemented, and the corresponding full-sky version is publicly available. Further details, including access to the implementation, can be found in Section \ref{sec:reproducible_research}.

\section{Generalisation of the Emulator to theoretical constraints}

The validation presented in this work relies on a target wavelet $\ell_1$-norm extracted from a simulation-generated weak lensing convergence map. However, the methodology developed here is not restricted to this specific validation case. In principle, the emulator allows for the substitution of the target values with any external input, including an arbitrary combination of a theoretical power spectrum $C_\ell$ and a corresponding set of $\ell_1$-norm values across wavelet scales. This adaptability enables the generation of statistically controlled convergence maps, making it possible to impose user-defined constraints without explicit dependence on simulation-based training data, that could come from purely theoretical grounds.

A notable advantage of this flexibility is the ability to incorporate stochasticity directly at the level of the input statistics so as to account for cosmic variance. In particular, any desired level of variance or non-Gaussian features arising from cosmic structure formation can be included by explicitly modulating the input target values. This approach allows for controlled sampling of cosmic variance, either by introducing uncertainty in the power spectrum $C_\ell$ or by perturbing the predicted $\ell_1$-norms within physically motivated bounds. Consequently, the emulator can be extended to produce ensembles of statistically consistent weak lensing maps that account for inherent fluctuations, thereby offering an alternative to expensive numerical simulations. One approach to achieve this could be to incorporate the cosmic variance at the level of the power spectrum only (see e.g \citet{moster2011cosmic}) and get the corresponding wavelet $\ell_1$-norm by only changing the power spectrum and keeping the scaled cumulants fixed \citet{2016MNRAS.460.1598C}.

\begin{figure}
    \centering
    \includegraphics[width=\linewidth]{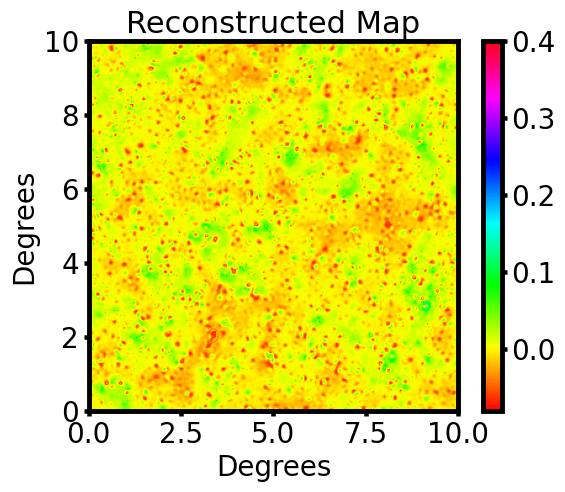}
    \caption{Emulated convergence map produced by our pipeline, \vilasini{with pixel size of $2\, \mathrm{arcmin}$,} generated by applying constraints on the first three scales for which the theory based wavelet $\ell_1$-norm is obtained. The x- and y-axes represent the map's range in degrees, with pixel values indicated by the colour bar.}
    \label{fig: target emulated theory}
\end{figure}

A fundamental limitation for using the theoretical prediction to generate maps, is its dependence on scale. The derivation constrains the statistical properties of the wavelet $\ell_1$-norm only up to mildly non-linear scales as was discussed before. However, for scales smaller than the constrained range, the generated map may not accurately reproduce the statistical properties of the simulation.

\begin{figure*}
    \centering
    \includegraphics[width=\textwidth]{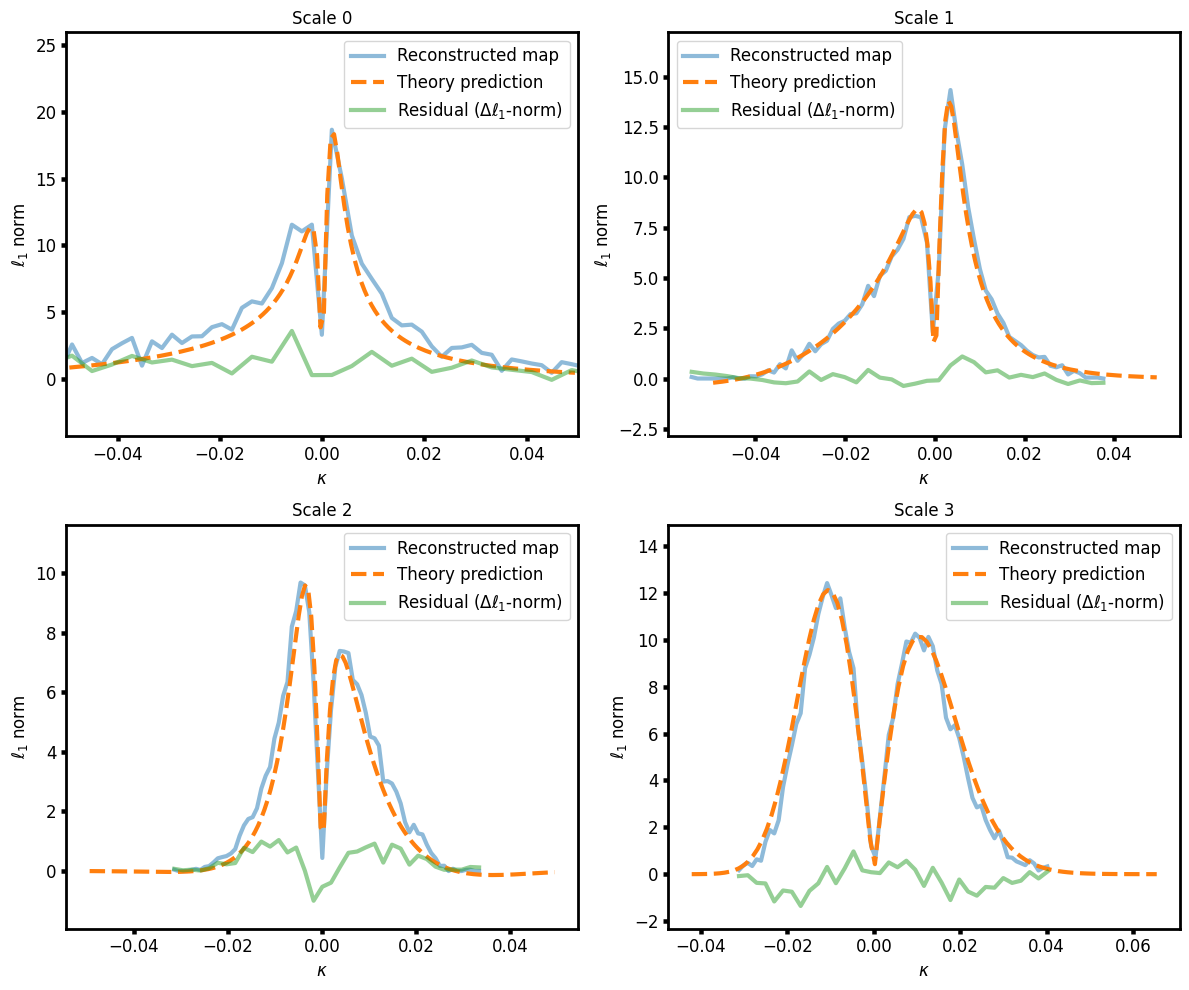}
    \caption{Comparison of the wavelet $\ell_1$-norm of the generated map (blue solid line) with the theoretical predictions (\vilasini{orange} dashed line) \vilasini{across different scales ranging from 2 to 8 $\mathrm{arcmin}$, shown from the top-left to the bottom-right panel, with the final panel representing the coarse scale at 8 $\mathrm{arcmin}$.} The residuals (\vilasini{green} solid line) between the target and emulated wavelet $\ell_1$-norms are also shown, indicating the accuracy of the emulation process in reproducing the expected distributions. The residuals consistently remain at a sub-percent level, demonstrating the precision of the reconstruction and the effectiveness of the algorithm. The small oscillations in the residuals are attributed to the finite patch size used in the analysis.}
    \label{fig: theory l1norm}
\end{figure*}

In Figure \ref{fig: target emulated theory}, we present an example of a map obtained by applying the constraints from the theory predictions at the first 3 scales (which includes the three wavelet coefficients and one coarse scale coefficient). We also compare the wavelet $\ell_1$-norm of the generated map with the theoretical predictions at the constrained scales, as shown in Figure \ref{fig: theory l1norm}. From the figure, it is evident that the generated map successfully reproduces the theoretical $\ell_1$-norm values with high accuracy. 

\section{Reproducible research}
\label{sec:reproducible_research}
The code GOLCONDA (Generative mOd-eLing of Convergence maps based ON predicteD one-point stAtistics) is made available in the spirit of open research. It enables the reproduction of the plots presented in this paper and can be accessed at \href{https://github.com/vilasinits/GOLCONDA}{GitHub}.

\section{Conclusions}
\label{sec:conclusions}

In this study, we developed and benchmarked a novel map emulation algorithm leveraging wavelet-based approaches, such as the tophat filter, to generate high-fidelity convergence maps with enhanced accuracy and computational efficiency. Our method demonstrates the ability to emulate maps of 500 pixels per side in approximately one minute, even when restricted to a single CPU core and without parallelisation or Just-In-Time  compilation. This performance underscores the practicality of our approach compared to traditional methods that typically require substantial computational resources. The pipeline's efficiency, coupled with its statistical rigor, provides a strong foundation for generating realistic cosmological maps in various applications.

A key focus of our work was the preservation and accurate reproduction of higher-order statistics, including variance, skewness, and kurtosis, at different scales and across entire maps. By conducting a thorough benchmarking analysis with top-hat filter, we demonstrated that our emulated maps closely approximate the statistical properties of target datasets, such as those derived from SLICS simulations. This capability to capture essential features of the underlying data makes our method particularly relevant for applications requiring precise modeling of small-scale structures and non-Gaussian features in cosmological fields.

Our results illustrate that the pipeline’s statistical alignment extends beyond basic moments, with a robust performance in reproducing higher-order metrics such as the cumulants across various smoothing scales apart from those used for constraining. This accuracy in representing complex structures suggests that our method provides a meaningful alternative to existing techniques, particularly log-normal map generation, which may not fully encapsulate higher-order dependencies and non-linear structures found in observed data.

Furthermore, the adaptability of our method offers promising opportunities for extensions and practical applications. The flexibility of the emulator enables the substitution of target values with any external input, including theoretical power spectra $C_\ell$ and corresponding sets of $\ell_1$-norm values across wavelet scales. This adaptability allows for the generation of statistically controlled convergence maps with user-defined constraints, without explicit reliance on simulation-based training data. Additionally, the ability to incorporate stochasticity at the input level provides a means to model cosmic variance explicitly, extending the emulator's capability to produce ensembles of statistically consistent weak lensing maps that account for inherent fluctuations.

\vilasini{While the present implementation of our algorithm is designed to reproduce specific one-point and two-point statistics—namely, the wavelet $\ell_1$-norm and the power spectrum—under idealised conditions, its modular structure allows for considerable flexibility. In particular, the framework can naturally be extended to incorporate observational systematics and physical effects. For instance, if theoretical predictions or simulation-based estimates of the wavelet $\ell_1$-norm and power spectrum are available for convergence fields affected by intrinsic alignments or baryonic feedback, these can be directly used as constraints in the emulator. This enables the generation of convergence maps that encode the impact of such effects without requiring explicit modeling of the underlying processes. Moreover, although systematics such as shear calibration errors and photometric redshift uncertainties typically arise earlier in the weak lensing analysis pipeline, our algorithm’s outputs can serve as inputs to downstream inference frameworks that incorporate these sources of uncertainty. Alternatively, the generated maps can be post-processed to simulate survey-specific features, including shape noise, masking, and selection biases. In summary, while this work has focused on validating the method in idealized scenarios, the algorithm’s design lends itself to future extensions that incorporate realistic systematics and survey conditions, either through modified input constraints or post-processing.}

Given these demonstrated capabilities, our wavelet-based emulation approach emerges as a robust and computationally efficient tool, potentially replacing or complementing lognormal map generation techniques. By accurately preserving both higher-order statistics and fine-scale structures, our method offers a more complete representation of complex cosmological fields. Future developments may focus on optimising the algorithm through parallelisation and JIT techniques, expanding its usability and applicability across a wider range of datasets and scales. Furthermore, this method has potential applications in inpainting. Inpainting techniques aim to reconstruct missing data in an image or signal by estimating the absent values from the available information. Various approaches, such as sparse inpainting using the Discrete Cosine Transform \citep{starck_image_2005,starck_morphological_2005}, sparse recovery techniques \citep{starck:pires08,lanusse16}, and Gaussian constrained realisations \citep{zaroubi95,starck:jeffrey18}, have been widely explored in the literature. The algorithm presented in this work could also prove useful for inpainting. While this exploration lies beyond the scope of this paper, assessing the effectiveness of this method compared to existing inpainting techniques remains an intriguing avenue for future research.

In conclusion, this work sets a foundation for efficient, accurate, and adaptable map emulation strategies, offering significant potential for advancing cosmological data analysis, large-scale simulations, and other applications where statistical fidelity and computational efficiency are paramount.

\begin{acknowledgements}
     This work was supported by the TITAN ERA Chair project (contract no. 101086741) within the Horizon Europe Framework Program of the European Commission, and the  Agence Nationale de la Recherche (ANR-22-CE31-0014-01 TOSCA).
     We would like to thank Joachim Harnois-Déraps for making public the SLICS mock data, which can be found at \url{http://slics.roe.ac.uk/}. 
     We would also like to thank Takahashi and collaborators for making public the TAKAHASHI simulation suite, which can be found at \url{http://cosmo.phys.hirosaki-u.ac.jp/takahasi/allsky_raytracing/}.
     The authors thank Andreas Tersenov, Cora Uhlemann and Alexandre Barthelemy for useful discussions. 
\end{acknowledgements}

\bibliographystyle{aa}
\bibliography{bibliography}

\appendix
\section{Algorithm in detail}
\label{app:projections}
\subsection{$\ell_1$-Norm Matching Wavelet Coefficient Adjustment Algorithm}

We are using the $\ell_1$-Norm Matching Wavelet Coefficient Adjustment Algorithm ($\ell_1$-WCA), which is designed to adjust wavelet coefficients at various scales to achieve specified $\ell_1$ norms within predefined histogram bins. This algorithm iteratively modifies the wavelet coefficients in each bin, separately adjusting positive and negative values to meet target $\ell_1$ norms while ensuring the values remain within set boundaries. We believe it could be useful in applications where specific statistical properties of wavelet-transformed data must be preserved, providing a flexible method to enforce norm constraints without disrupting the underlying data distribution.

Given a target $L_{1,t}$, the wavelet coefficient's $\ell_1$ can be adjusted to match the target as follows:

First the total adjustment $\Delta_{total}$, is obtained as the difference between the target and the current wavelet $\ell_1$-norm. 

\begin{align}
\Delta_{\text{total}} &= L_{1,t}^{\theta_j} - \sum_{i=1}^{N_{\text{bin}}} |w_{j,i}|,  
\end{align}

where $ L_{1,t}^{\theta_j} $ represents the target $\ell_1$-norm at scale $j$, and the summation computes the current $\ell_1$-norm of the wavelet coefficients. This is used to get the adjustment per positive coefficient as:

\begin{align}
\Delta_p &= \frac{\Delta_{\text{total}}}{N_p},  
\end{align}

where $N_p$ is the number of positive coefficients. The adjusted positive coefficients are then given by:

\begin{align}
w_{j,i} &= \max(w_{j,i} + \Delta_p, 0), \quad \forall i \text{ such that } w_{j,i} \geq 0.
\end{align}

After this update, the new $\ell_1$ is computed as:

\begin{align}
L_{1,\text{updated}} &= \sum_{i=1}^{N_{\text{bin}}} |w_{j,i}|.
\end{align}

The remaining adjustment to the target norm is obtained as:

\begin{align}
\Delta_{\text{residual}} &= L_{1,t}^{\theta_j} - L_{1,\text{updated}}.
\end{align}

The residual is distributed amongst the negative coefficients and updated as:

\begin{align}
w_{j,i} &= \min(w_{j,i} - \Delta_n, 0), \quad \forall i \text{ such that } w_{j,i} < 0.
\end{align}

This now gives a wavelet coefficient with the constrained $\ell_1$-norm matching the target.

The corresponding algorithm is given in Algorithm \ref{algo: L1WCA}.

\begin{algorithm}
\label{algo: L1WCA}

\caption{$\ell_1$-Norm Matching Wavelet Coefficient Adjustment ($\ell_1$-WCA)}
\label{algo:l1_wca}

\textbf{Input:} 
\begin{itemize}
    \item Wavelet coefficient set at scale $j$: $w_j$
    \item Target $\ell_1$-norm: $L_{1,t}^{\theta_j}$
    \item Number of histogram bins: $N_{\text{bin}}$
\end{itemize}

\textbf{Output:} Adjusted wavelet coefficients $w_j$

\begin{algorithmic}[1]
    \STATE Compute current $\ell_1$-norm: $L_{1,c} = \sum_{i=1}^{N_{\text{bin}}} |w_{j,i}|$
    \STATE Compute total adjustment: $\Delta_{\text{total}} = L_{1,t}^{\theta_j} - L_{1,c}$
    \STATE Compute adjustment per positive coefficient: $\Delta_p = \frac{\Delta_{\text{total}}}{N_p}$, where $N_p$ is the number of positive coefficients
    \FOR{each coefficient $w_{j,i}$}
        \IF{$w_{j,i} \geq 0$}
            \STATE $w_{j,i} \gets \max(w_{j,i} + \Delta_p, 0)$
        \ENDIF
    \ENDFOR
    \STATE Compute updated $\ell_1$-norm: $L_{1,\text{updated}} = \sum_{i=1}^{N_{\text{bin}}} |w_{j,i}|$
    \STATE Compute residual adjustment: $\Delta_{\text{residual}} = L_{1,t}^{\theta_j} - L_{1,\text{updated}}$
    \STATE Compute adjustment per negative coefficient: $\Delta_n = \frac{\Delta_{\text{residual}}}{N_n}$, where $N_n$ is the number of negative coefficients
    \FOR{each coefficient $w_{j,i}$}
        \IF{$w_{j,i} < 0$}
            \STATE $w_{j,i} \gets \min(w_{j,i} - \Delta_n, 0)$
        \ENDIF
    \ENDFOR
    \FOR{each coefficient $w_{j,i}$}
        \STATE Enforce boundaries: $w_{j,i} \gets \max(\min(w_{j,i}, \text{BoundMax}), \text{BoundMin})$
    \ENDFOR
    \RETURN $w_j$
\end{algorithmic}
\end{algorithm}

\subsection{Algorithm for getting the correct powerspectra}

To generate an image with the desired target power spectrum $(\mathrm{Cls})$ we perform Fourier amplitude rescaling. This process adjusts the Fourier amplitudes of the image to match the target power spectrum while preserving the original phase information to maintain spatial structure as shown in Algorithm \ref{algo:fourier_amplitude_correction}.

\begin{algorithm}
\caption{Fourier Amplitude Rescaling to Match Target Power Spectrum}
\label{algo:fourier_amplitude_correction}

\textbf{Input:} 
\begin{itemize}
    \item Input map $\kappa$
    \item Target power spectrum $P_t$
\end{itemize}

\textbf{Output:} Adjusted map $\kappa'$

    \begin{algorithmic}[1]
        \STATE Compute the Fourier transform of the input map: $\hat{\kappa}(\mathbf{k}) = \mathcal{F}[\kappa]$
        \STATE Compute the radial power spectrum of the input map: $P_{\kappa} (k) = |\hat{\kappa} (\mathbf{k})|^2$
        \STATE Compute the rescaling factor: $R(k) = \sqrt{\frac{P_t(k)}{P_{\kappa} (k)}}$
        \STATE Rescale the Fourier amplitudes: $\hat{\kappa}'(\mathbf{k}) = R(k) \cdot \hat{\kappa}(\mathbf{k})$
        \STATE Perform the inverse Fourier transform to obtain the adjusted map: $\kappa' = \mathcal{F}^{-1}[\hat{\kappa}'(\mathbf{k})]$
        \RETURN $\kappa'$
    \end{algorithmic}
\end{algorithm}

\section{Evaluation of higher order statistics}
\label{app:HOS}
\vilasini{We conducted a detailed comparison of the emulated convergence maps with the target maps, focusing on the one-point Probability Distribution Function (PDF) and peak counts.}
In addition to the main analysis presented in the main text, we have also conducted a detailed comparison of the emulated convergence maps with the target maps using other high-order statistics. We have especially focused on the one-point Probability Distribution Function (PDF) and peak counts as detailed below. 

\subsection{Probability Distribution Function (PDF)}

\begin{figure*}
    \centering
    \includegraphics[scale=0.4]{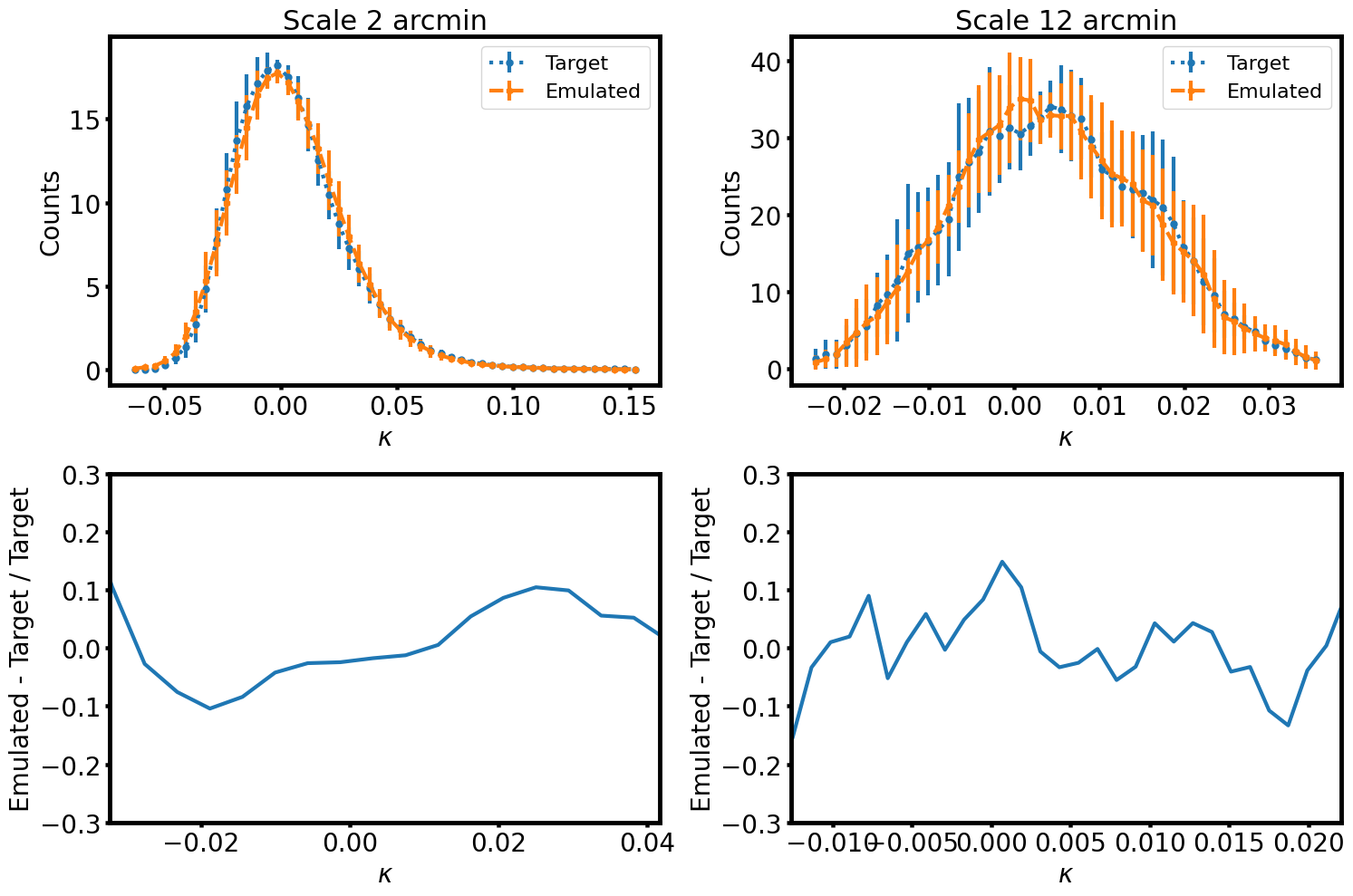}
    \caption{Top panels: One-point PDF of the target and emulated maps at two smoothing scales (2 and 12 arcmin), computed using a circular top-hat filter. Bottom panels: Residuals between emulated and target PDFs.}
    \label{fig:pdf_scales}
\end{figure*}

\vilasini{Figure~\ref{fig:pdf_scales} shows the one-point PDF of the convergence field at two different smoothing scales (2 arcmin and 12 arcmin), obtained using a circular top-hat filter. These filters are distinct from the wavelet-like difference-of-top hats used in defining the $\ell_1$-norm constraint. The top panels show the PDFs for both the target and emulated maps, while the bottom panels display the residuals. The results indicate agreement at the $\sim10\%$ level across most of the distribution, with slight discrepancies in the extreme tails.}
Let us note that for intermediate to large smoothing scales (e.g 12 arcmin here), the difference is completely within the statistical error bars of our measurements, hence no significant departure is observed in this regime. For smaller smoothing scales, the deviation is more significant but this is in a regime where non-linearities start to be quite important (hence less interesting for cosmological analysis if systematics can not be controlled) and resolution effects might also play a significant role.

\subsection{Peak Count Statistics}

\begin{figure*}
    \centering
    \includegraphics[scale=0.5]{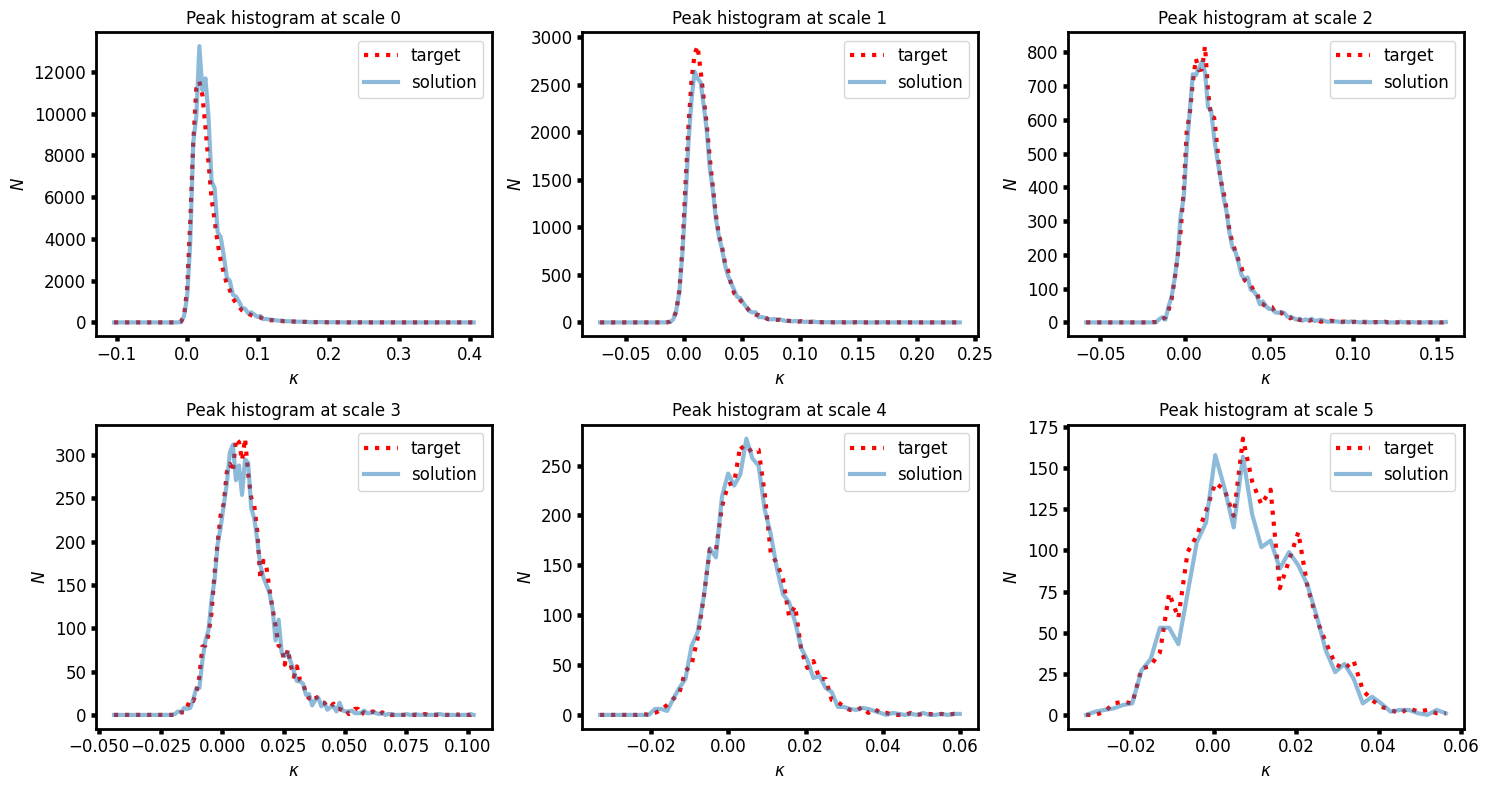}
    \caption{Peak count distribution at different wavelet scales as labeled for the target (dotted red line) and emulated maps (blue solid line). Emulated maps reproduce the peak counts of the target with high fidelity at small to intermediate scales.}
    \label{fig:peak_count}
\end{figure*}

\vilasini{In Figure~\ref{fig:peak_count}, we show the peak count distribution at various wavelet scales. Peaks are defined as local maxima in the map, and the histograms are shown separately for the target and emulated maps. The agreement is very good for all scales, with some fluctuation at the largest scales (4 and 5) mostly driven by a lack of statistics.
}

\subsection{Implications for Higher-Order Moments}

\vilasini{Although the algorithm is optimised only to match the power spectrum and wavelet $\ell_1$-norm, the consistency seen in the PDFs and peak counts especially in the tails indicates that higher-order moments are approximately preserved. These statistics, especially the peak counts, are sensitive to non-Gaussian features and implicitly reflect the presence of higher-order correlations beyond the fourth moment. However, we emphasise that our method does not guarantee fidelity in any specific higher-order moment. The observed agreement suggests the approach generalises well and captures broader statistical features, but also leaves room for future improvements that target additional moments explicitly.}

\end{document}